\documentclass{vgtc}                          

\ifpdf
  \pdfoutput=1\relax                   
  \usepackage{graphicx}                
  \DeclareGraphicsExtensions{.pdf,.png,.jpg,.jpeg} 
\else
  \usepackage{graphicx}                
  \DeclareGraphicsExtensions{.eps}     
\fi%

\graphicspath{{figures/}{pictures/}{images/}{./}} 

\usepackage{microtype}                 
\PassOptionsToPackage{warn}{textcomp}  
\usepackage{textcomp}                  
\usepackage{mathptmx}                  
\usepackage{times}                     
\usepackage{cite}                      
\usepackage{tabu}                      
\usepackage{booktabs}                  
\usepackage{multirow}

\usepackage{amsmath, amssymb}
\usepackage{type1cm}
\usepackage{enumitem}
\usepackage[whole]{bxcjkjatype}
\usepackage{algorithm}
\usepackage{algpseudocode}
\usepackage{here}

\onlineid{0}

\vgtccategory{Research}

\vgtcinsertpkg

\DeclareMathAlphabet\mathbfcal{OMS}{cmsy}{b}{n}
\DeclareMathAlphabet\mathbfit{OML}{cmm}{b}{it}

\title{Information Entropy-based Camera Path Estimation for In-Situ Visualization}

\author{Ken Iwata\thanks{e-mail: 228x202x@stu.kobe-u.ac.jp}\\ %
        \scriptsize Kobe University %
\and Naohisa Sakamoto\thanks{e-mail: naohisa.sakamoto@people.kobe-u.ac.jp}\\ 
     \scriptsize Kobe University %
\and Jorji Nonaka\thanks{e-mail: jorji@riken.jp}\\ %
     \scriptsize RIKEN R-CCS
\and Chongke Bi\thanks{e-mail: bichongke@tju.edu.cn}\\ %
     \scriptsize Tianjin University}

\teaser{
  \centering
  \includegraphics[width=\linewidth]{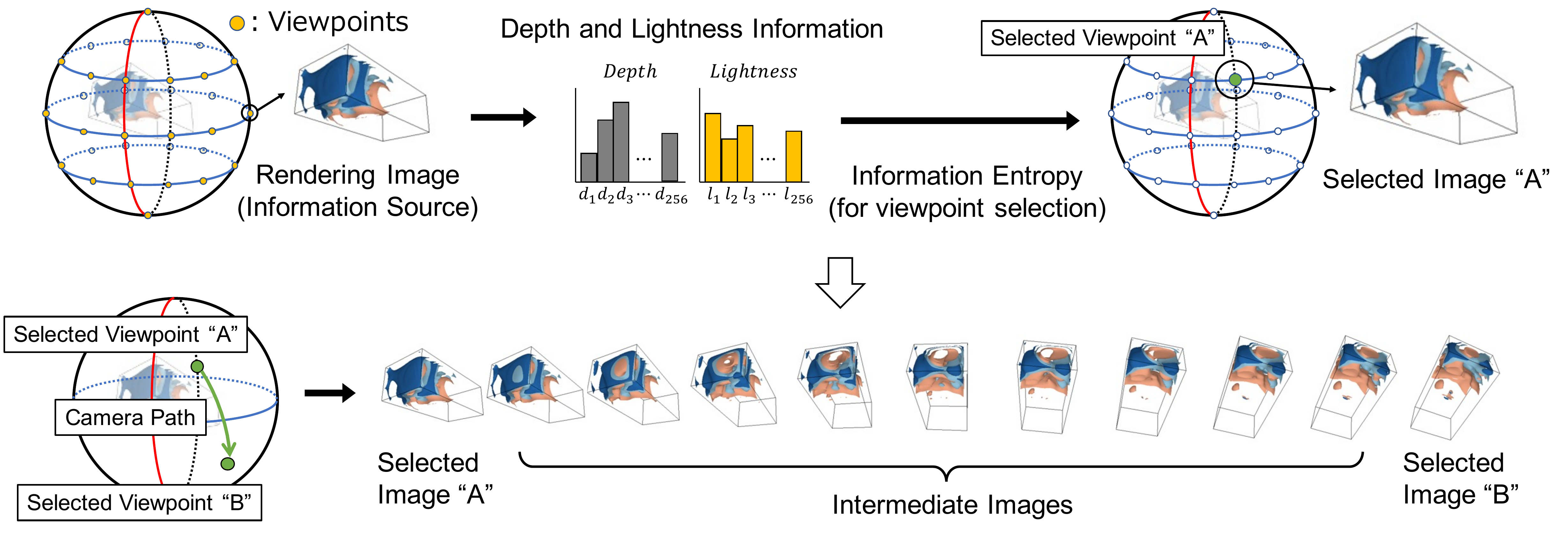}
  \caption{Depth and lightness entropy-based viewpoint selection and camera path estimation for generating a smooth video, with as much information as possible, to assist the rapid understanding of the underlying simulation phenomena.}
  \label{fig:teaser}
}

\abstract{
In-situ processing has widely been recognized as an effective approach for the visualization and analysis of large-scale simulation outputs from modern HPC systems. One of the most common approaches for batch-based in-situ visualization is the image- or video-based approach. In this kind of approach, a large number of rendered images are generated from different viewpoints at each time step \color{black} and has proven useful for detailed analysis of the main simulation results. However, during test runs and model calibration runs before the main simulation run, a quick overview might be sufficient and useful. \color{black} In this work, we focused on selecting the viewpoints which provide as much information as possible by using information entropy to maximize the subsequent visual analysis task. However, by simply following the selected viewpoints at each of the visualization time steps will probably lead to a rapidly changing video, which can impact the understanding. Therefore, we have also worked on an efficient camera path estimation approach for connecting selected viewpoints, at regular intervals, to generate a smooth video. This resulting video is expected to assist in rapid understanding of the underlying simulation phenomena and can be helpful to narrow down the temporal region of interest to minimize the turnaround time during detailed visual exploration via image- or video-based visual analysis \color{black} of the main simulation run. \color{black}We implemented and evaluated the proposed approach using the OpenFOAM CFD application, on an x86-based Server and an ARM A64FX-based supercomputer (Fugaku), and we obtained positive evaluations from domain scientists.
}

\CCScatlist{
  \CCScatTwelve{Human-centered computing}{Visualization}{Visualization systems and tools}{Visualization toolkits}
}

\begin{document}

\firstsection{Introduction}

\maketitle






High-end \color{black}high performance computing (HPC) \color{black}systems have continuously become more and more capable with higher computational capacity with every new system replacement. This was the case for the replacement of the \textit{K computer} to the \textit{supercomputer Fugaku} at the RIKEN R-CCS. The increased number of CPUs and computational cores have been applied for {\it Capability Computing} to tackle even larger numerical simulations with higher spatio-temporal resolutions. In addition, this also has been used for {\it Capacity Computing} to handle an even larger number of parameters and members during parametric sweep and ensemble simulations. On the other hand, this proportionately generates even larger simulation outputs, thus, making the visualization and analysis tasks even more challenging. As a result, the importance of in-situ visualization and analysis has continuously become even more evident.

A variety of approaches have already been proposed and applied for the in-situ visualization and analysis as discussed in~\cite{childs2020terminology}. We can also verify that there are also a variety of existing applications and libraries for realizing in-situ visualization and analysis. However, since in-situ processing is executed simultaneously with the simulation, it becomes highly important to collaborate with the domain scientists \color{black}. We have been working with domain scientists working with computational fluid dynamics (CFD) simulation of the sound generation mechanisms~\cite{yoshinaga2018experimental}, and we already worked on an adaptive in situ time-step sampling approach~\cite{yamaoka2019situ}. In this work, we have used the same OpenFOAM CFD application and simulation model and obtained assistance from them for necessary technical feedback during the developments. \color{black} 



Probably the most widely used image-based in-situ visualization approach is ParaView Cinema~\cite{ahrens2014image}. In that approach, a large set of pre-computed images are generated in-situ on the HPC system side for the interactive post-hoc visual exploration on a local machine such as desktop PC and laptop. There is also an image-based in-situ visualization approach that generates a set of images from omnidirectional camera positions~\cite{ kageyama2014approach}, and its extension for video-based in-situ visualization~\cite{kageyama20204dstreetview}. \color{black}These image- or video-based in-situ visualization approaches have proven useful for detailed analysis of the main simulation results. In this work, we have focused on rapid understanding of the underlying simulation during test runs and model calibration runs before the main simulation run. For this purpose, we focused on selecting the most appropriate viewpoints, based on information entropy, at regular time intervals of the simulation in order to obtain as much information as possible trying to facilitate the rapid understanding of such kinds of simulations. \color{black}



\section{Related Work}

There is an extensive work that culminated in the creation of a classification and terminology for the in-situ visualization approaches~\cite{childs2020terminology}. Here, we will only focus on related works for realizing tightly coupled in-situ visualization, and techniques for selecting time steps and viewpoints that can be used for minimizing the amount of images for the image- or video-based in-situ visualization. VTK-based ParaView and VisIt are probably the most widely used visualization application for large data visualization. Both applications provide in-situ visualization APIs, {\it ParaView Catalyst}~\cite{ayachit2015paraview} and {\it VisIt LibSim}~\cite{kuhlen2011parallel}, for integrating to the simulation code. In a batch-based in-situ visualization, a large amount of images can be generated for assisting the post-hoc visualization~\cite{kageyama2014approach}. To facilitate this post-hoc visual analysis, Ahrens~et~al.~\cite{ahrens2014image} proposed an image-based approach for the in-situ visualization and analysis, and was implemented as {\it Paraview Cinema}. In this approach a large set of images are generated in-situ, and a custom visualization application is used, on the local machine, to perform interactive visual analysis by automatically switching between the generated set of images. Similar to this approach, Kageyama~et~al.~\cite{kageyama20204dstreetview} proposed a video-based approach by generating an omnidirectional animated video, from the set of in-situ generated images, which are explorable from a custom visualization application. Although these approaches have proven efficient, most of the generated images may have small or even no contribution to the visual analysis, thus it may be unnecessarily increasing the time spent on the post-hoc visual analysis task.


An approach to minimize the aforementioned amount of generated images is the selection of the most valuable time steps for rendering the images. For this purpose, Ling~et~al.~\cite{ling2017using} proposed a method to estimate the probability density function of the simulation field, at each time step, by using the kernel density estimation. They also applied machine learning for extracting feature quantity from the obtained estimation, and detected potentially valuable time steps where an important phenomena may occur. However, this method can cause false detections depending on the high correlation among the physical quantities on the simulation field as mentioned by the authors. Yamaoka~et~al.~\cite{yamaoka2019situ} extend the aforementioned work, and proposed an adaptive time sampling method for in-situ visualization. In this method, kernel density function and Kullback–Leibler divergence is applied to estimate the amount of change on the simulation field. The sampling intervals are adaptively changed according to the estimated amount of change in the simulation. 




Another approach for reducing the amount of images is the selection of viewpoints for generating the images. For this purpose, Kamada~et~al.~\cite{kamada1988simple} considered the viewpoints capable of minimizing the number of degenerated face as being the optimal viewpoints. However, they did not extend their work for the case when there exist multiple viewpoints with the same number of degenerated faces. Barral~et~al.~\cite{barraly2000scene} solved this problem by adding the projected area as a weight to the number of degenerated faces. However, there still remains a problem on how to properly set these weights. Vazquez~et~al.~\cite{vazquez2001viewpoint} proposed a method to select the optimal viewpoint defined by the viewpoint entropy based on the information entropy. Since this method cannot handle the movement of viewpoints, the authors improved the viewpoint entropy and applied it to molecular objects~\cite{vazquez2002viewpoint} as well as to image-based modeling~\cite{vazquez2003automatic}. Page~et~al.~\cite{page2003shape} proposed a method to analyze the object shape by calculating the entropy for the silhouette and surface curvature of the model. Polonsky~et~al.~\cite {polonsky2005s} discussed evaluation indices for the viewpoint selection, and concluded that none of them could make the best choice in any situation. However, they also said that by improving each of these indices, it will become possible to make a better choice by using a combination of them. 


Secord~et~al.~\cite{secord2011perceptual} proposed some evaluation indices for the viewpoint selection, and showed that optimal  viewpoints can be selected by combining these metrics. Takahashi~et~al.~\cite{takahashi2005viewpoint} proposed a method for estimating the optimal viewpoint for volume data by using information entropy. Bordoloi~et~al.~\cite{bordoloi2005viewpoint} proposed an information entropy-based evaluation metric for the viewpoints during volume rendering by using the transfer function, data distribution, and voxel visibility information. Zhang~et~al.~\cite{zhang2010optimal} also proposed an evaluation metric for volume rendering based on the opacity, brightness, and structural features. Ji~et~al.~\cite{ji2006viewpoint} proposed a method to find the optimal time-varying views by using the viewpoint selection method to maximize the amount of information for time-series volume data. They showed that it is possible to create an animation with the largest amounts of information. This was realized by searching for a movement route with the largest amounts of information using dynamic programming.  Marsaglia~et~al.~\cite{marsaglia2021entropy} proposed a viewpoint quality evaluation metric based on \color{black}information entropy involving the visible field data, depth, and shading values belonging to each of the pixels in the image. In another work, they also  utilized a trigger-based approach in combination with information entropy to determine when to search for a new camera position as a simulation evolves~\cite{marsaglia2022camera}. Our work was inspired in their viewpoint quality evaluation metric, which we extended with the lightness information \color{black} for evaluating the viewpoint quality. We will detail the methodology behind our proposed method in the next section.


\section{Methodology}


\subsection{Overview}

In this work, we focused on a viewpoint selection approach, based on information entropy, and on a camera path estimation approach, based on quaternion interpolation. The viewpoints selected at regular intervals are used as markers to estimate the smooth camera path. Following are the necessary requirements to meet this goal:

    \begin{description}
        \item [R1.] Images from the selected viewpoints should capture important phenomenon from the underlying simulation.
        \item [R2.] The resulting video generated from the rendered outputs should be smooth for post-hoc analysis.
    \end{description}

Below is the adopted approach to satisfy the aforementioned requirements, and they are divided into the following three parts:

    \begin{description}
        \item[A. Viewpoint evaluation] \mbox{}\\
        The viewpoint quality will be evaluated using information entropy and will be used to select the most appropriate image for each evaluated time step. Only images from the selected viewpoints will be output ({\bf R1}).
        \item[B. Camera path estimation] \mbox{}\\
        The camera path connecting these selected viewpoints will be  estimated, and the rendered images through this camera path will also be output as intermediate images ({\bf R2}).
        \item[C. Video generation] \mbox{}\\
        At the end of the simulation, these output images will be sequentially concatenated to produce a video ({\bf R2}).
    \end{description}

\begin{figure*}[t!]
    \centering
    \includegraphics[width=0.85\linewidth]{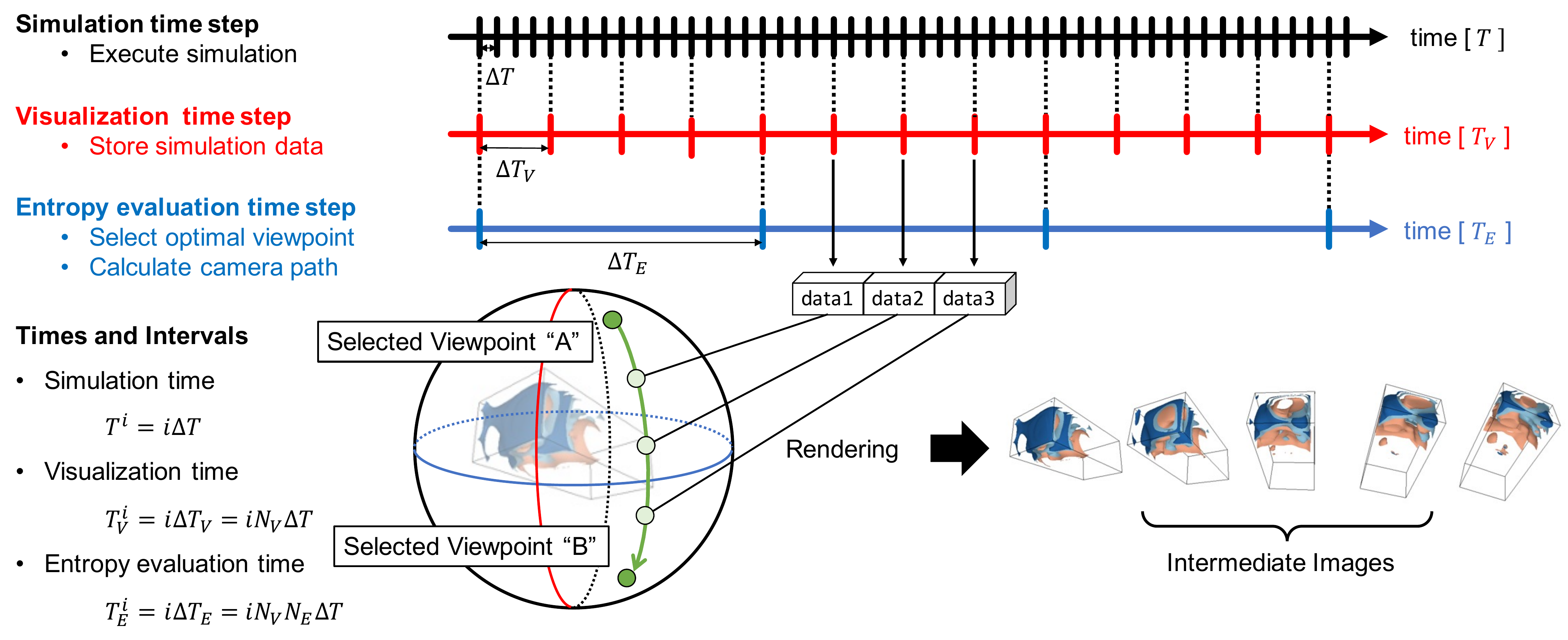}
    \caption{Different time step intervals for the simulation, visualization and entropy evaluation.}
    \label{fig:timesteps}
\end{figure*}

Regarding part A, the simulation state usually does not often significantly change within a single simulation time step. Therefore, there is usually no need to visualize at every simulation time step, and the visualization can be performed at every set of simulation time steps. In the same manner, the viewpoint evaluation for viewpoint selection will be performed at every set of visualization time steps to satisfy {\bf R1}. In this paper, as shown in  Fig.~\Ref{fig:timesteps}, we use $T$ to represent the simulation time step, $T_{V}$ to represent the visualization time step, and $T_E$ to represent the entropy evaluation time step, $\Delta T$ to represent the simulation time step interval, $\Delta T_{V}$ to represent the visualization time step interval, and $\Delta T_{E}$ to represent the entropy evaluation time step interval. The elapsed time for the $i$th simulation time step $T^{i}$ can be expressed as $T^{i} = i \Delta T$. In the same manner, the visualization time step and entropy evaluation time steps can be expressed respectively as $T_{V}^{i} = i \Delta T_{V}$ and $T_{E} ^ {i} = i \Delta T_{E}$. Considering that the simulation is performed $N_{V}$ times for every visualization, and the visualization is performed $N_{E}$ times for every optimal viewpoint selection, then we can express these intervals as $\Delta T_{V} = N_ {V} \Delta T$ and $ \Delta T_{E} = N_{E} N_{V} \Delta T$.



Regarding part B, a camera path connecting viewpoints selected at every $\Delta T_{E}$ entropy evaluation time step will be estimated. Although visualization is not performed during the visualization time steps in between the entropy evaluation time steps, the simulation data for each $\Delta T_{V}$ visualization time step is stacked. From the obtained camera path, the rendered images at the intermediate visualization time steps will be output as the intermediate images for generating a smooth video, and this satisfies {\bf R2}. It is worth noting that it is also possible to generate the full set of images from the entire viewpoints for the detailed post-hoc analysis when necessary.


Regarding part C, the set of output images generated at each $\Delta T_{V}$ visualization time step will be joined sequentially to create a video file. For this purpose, we can use existing tools such as the well-known {\it FFmpeg} available to a variety of platforms. In the resulting video, the camera will automatically move and capture the important phenomenon, and this allows the {\bf R2} to be satisfied. Traditional approach requires the user to search for the best location, in the trial-and-error manner, to visually explore when searching for an important phenomenon during the simulation. However, by using the proposed method, this search for the best camera position may be alleviated and may facilitate narrowing down the spatio-temporal region of interest for the detailed visual analysis.

\subsection{Viewpoint Selection}
In this section, we will detail the utilized viewpoint selection approach. The evaluation of the viewpoints is based on information entropy. We used depth and lightness values from the rendered images for calculating the associated information entropy, that is, the depth entropy and lightness entropy.

\subsubsection{Information Entropy}



Information entropy used in this work can be defined as the expected value for the amount of information obtained from a certain information source~\cite{blahut1987principles}. The information entropy $H(X)$ from a source $X$ given by the set of probabilities $P(x_{1}), P(x_{2}), \cdots, P(x_{n})$ corresponding respectively to the set of information $x_{1}, x_{2}, \cdots, x_{n}$ can be represented as follows:

\begin{equation}
    H(X)=-\sum^{n}_{i=1} P(x_{i})\log P(x_{i}) 
\end{equation}

Here, when $P(x_{i}) = 0$, we will consider $P(x_{i}) \log P(x_{i}) = 0$. \color{black}Regarding the selection of the logarithm's base, the base influences the multiplication factor and, thus, is arbitrary. Base 2 is commonly used in information theory, and was used in this work. \color{black}Taking into consideration the probability distribution of the information source, the larger the information bias, the smaller the value of information entropy, and vice versa.

\subsubsection{Depth Entropy}
Depth entropy used in this work is based on the viewpoint quality evaluation metric proposed by Marsaglia~et~al.~\cite{marsaglia2021entropy}. The information entropy is calculated by considering the image as the source of information, and by using the depth values belonging to each of the pixels in the image. The depth values can vary in the range of $0 \sim 1$, and the closer the distance to the object, the smaller the value. The background portion in the image where there is no object is set to infinity and will have their depth values corresponding to 1.

\color{black}The depth values from all pixels of the rendered image are binned into 256 groups \color{black}$d_{0}, d_{1}, \cdots, d_{255}$ for creating a discrete probability distribution $D$, which will be used to calculate the information entropy. At this time, the background portion in the image is considered to have no information, and only the pixels with some information will be used in the calculation. By using the discrete probability distribution, the depth entropy $H_{d}$ can be calculated as follows:

\begin{equation}
    H_{d}=-\sum^{255}_{i=0} D(d_{i})\log_{2} D(d_{i})
\end{equation}

Here, $D(d_{i})$ corresponds to the probability for a given value, selected based on the probability distribution $D$, being $d_{i}$. When evaluating a viewpoint using depth entropy, the resulting value will be larger for images with large dispersion in the distribution of depth values. Therefore, the viewpoints of images showing objects with high undulations will be highly evaluated.

\subsubsection{Lightness Entropy}
\begin{figure}[hbt!]
    \centering
    \includegraphics[width=\linewidth]{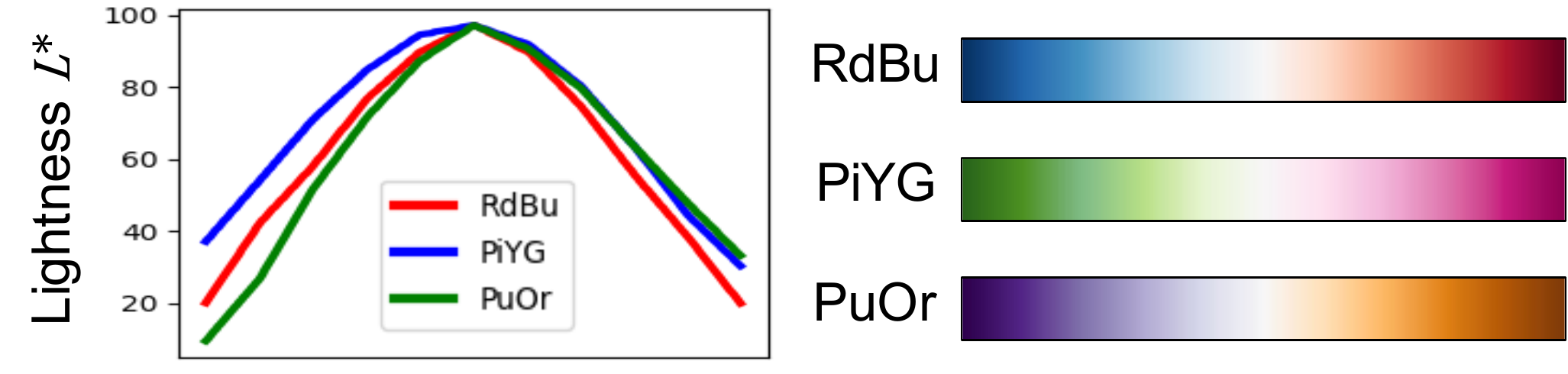}
    \caption{Lightness values for different diverging color maps.}
    \label{fig:diverging_color_map}
\end{figure}

In this work, in addition to the depth entropy, we propose the use of lightness entropy to also take into consideration the color information in the image. Diverging color maps, proposed by Moreland~\cite{moreland2009diverging}, have become prevalent in scientific visualization as the substitute for the traditional but problematic rainbow color map. Although the change in color values, such as RGB values, in a color map may follow different behavior depending on the color map, diverging color maps usually show similar behavior in the lightness values in CIELAB color space as shown in the Fig.~\ref{fig:diverging_color_map}. Therefore, lightness entropy is expected to work robustly for the diverging color maps. The proposed lightness entropy can be defined as an information entropy using the lightness values from the target image as the source of information. The lightness value is calculated from RGB values and can vary in the range of $0 \sim 100$. 

\color{black}The lightness values, as well as the depth values, obtained from all pixels of the rendered image, are binned into 256 groups \color{black}$l_{0}, l_{1}, \cdots, l_{255}$ for creating a discrete probability distribution $L$, which will be used to calculate the information entropy. At this time, the background portion in the image is considered to have no information, and only the pixels with some information will be used in the calculation. By using the discrete probability distribution, the lightness entropy $H_{l}$ can be calculated as follows:

\begin{equation}
    H_{l}=-\sum^{255}_{i=0} L(l_{i})\log_{2} L(l_{i})
\end{equation}

Here, $L(l_{i})$ corresponds to the probability for a given value, selected based on the probability distribution $L$, being $l_{i}$. When a viewpoint is evaluated using the lightness entropy, the resulting value will be larger for images with large dispersion in the distribution of lightness values. Therefore, the viewpoints of images with clear brightness and darkness will be highly evaluated.

\subsection{Path Estimation between Selected Viewpoints}

\begin{figure}[h]
    \centering
    \includegraphics[width=4.5cm]{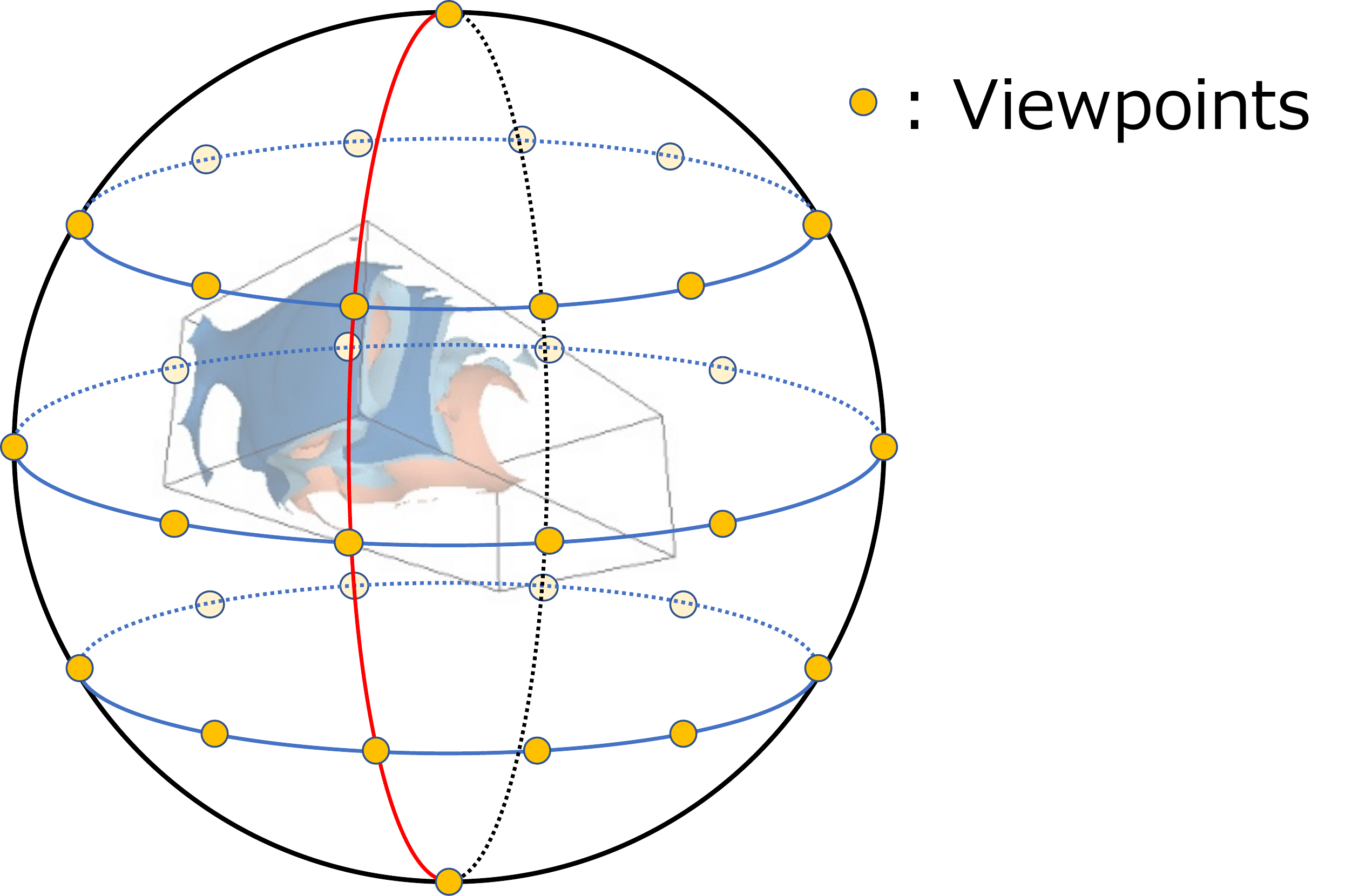}
    \caption{Viewpoint distribution over spherical surface.}
    \label{fig:spherical}
\end{figure}

In this section, we will detail the utilized camera path estimation between the viewpoints selected by using the depth and/or lightness entropy. In this work, we considered that the viewpoints are pre-arranged in a spherical surface as shown in Fig.~\ref{fig:spherical}, and the camera path from one viewpoint to another will move over this spherical surface. More specifically, the position and orientation of a given viewpoint will be represented as a quaternion, and the movement from one to another viewpoint will be obtained by using quaternion interpolation. In this work, we investigated two quaternion interpolation methods: spherical linear interpolation (SLERP) and spherical quadrangle interpolation (SQUAD). In the following subsections, we will explain about spherical linear interpolation and spherical quadrangle interpolation.

\subsubsection{Spherical Linear Interpolation (SLERP)}
SLERP is an abbreviation for spherical linear interpolation, and is a quaternion interpolation method for connecting two points over a sphere in the straight line direction, or the shortest path, as shown in Fig.~\ref{fig:interpolations}. SLERP-based interpolation from a quaternion $q_{A}$ to the quaternion $q_{B}$ can be calculated by using time $t \in [0, \, 1]$ as follows:

\begin{equation}
    slerp(q_{A},\,q_{B},\,t)=\frac{\sin{(1-t)\phi}}{\sin{\phi}}q_{A}+\frac{\sin{t\phi}}{\sin{\phi}}q_{B}
\end{equation}

Here, $\phi = \arccos {\langle q_{A}, \, q_{B} \rangle}$. In addition, in the case of $\langle q_{A}, \, q_{B} \rangle < 0$, the interpolation will be interpolated in the contrary direction over the sphere surface, that is by the longest path in the straight-line direction. To interpolate by the shortest path, then either $q_{A} $ or $ q_{B}$ should be replaced with a quaternion with same rotation but in the opposite direction. For instance, by replacing $q_{A}$ with $-q_{A}$.


\subsubsection{Spherical Quadrangle Interpolation (SQUAD)}
\begin{figure}[h]
    \centering
    \includegraphics[width=5.5cm]{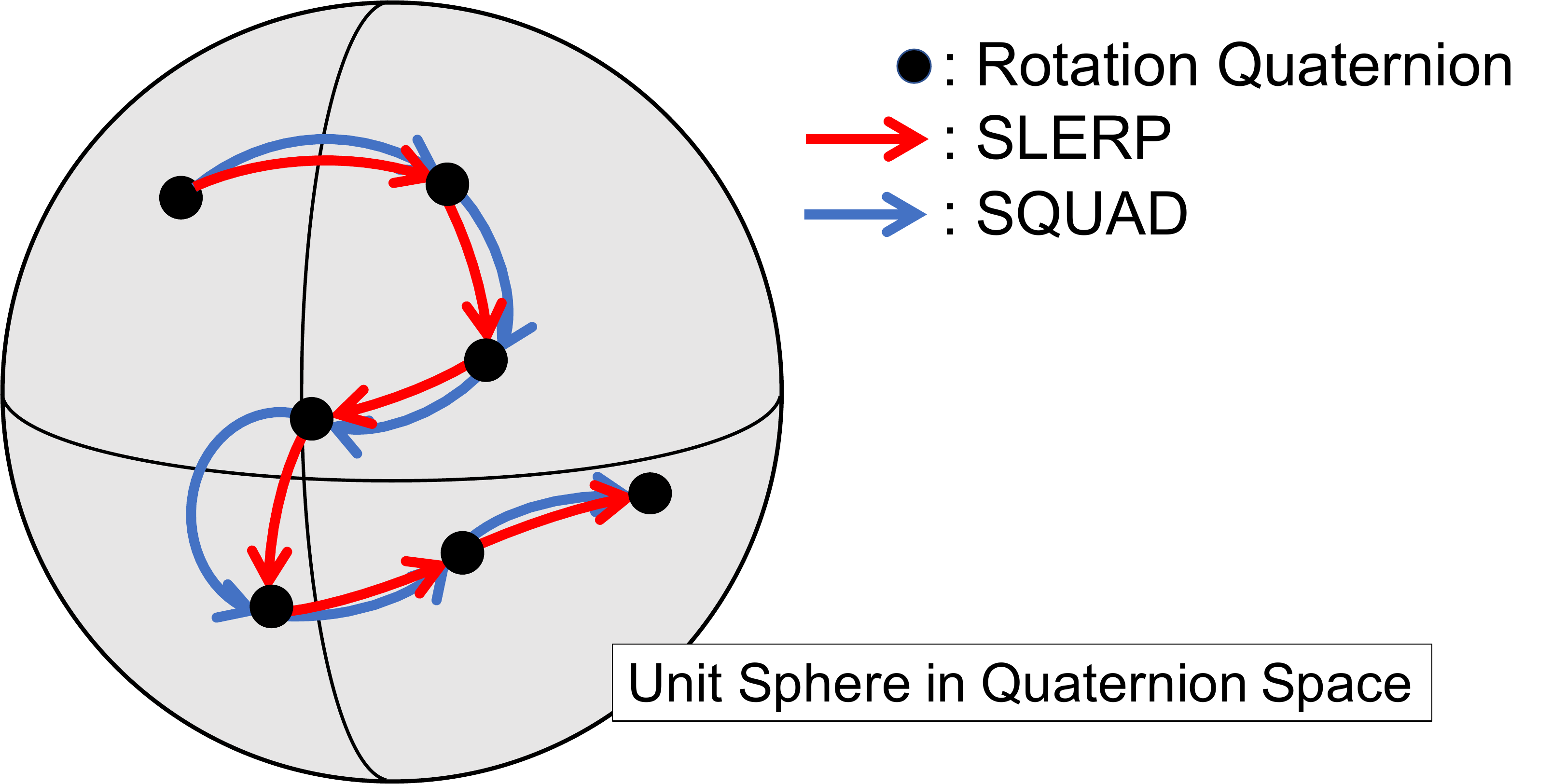}
    \caption{Comparison of SLERP and SQUAD interpolation methods.}
    \label{fig:interpolations}
\end{figure}

SQUAD is an abbreviation for spherical quadrangle interpolation, and is a quaternion interpolation method to connect multiple points in a smoothness way so that the derivatives are continuous in the neighborhood of the points (Fig.~\ref{fig:interpolations}). Considering a quaternion list $\{q_{1}, \, q_{2}, \, \dots \, q_{n}\}$, then the SQUAD-based interpolation from $q_{i}$ to $q_{i + 1}$ can be calculated by using the time $t \in [0, \, 1]$ as follows:

\begin{eqnarray}
    &squad(q_{i},\,q_{i+1},\,a_{i},\,a_{i+1},\,t)\notag\\&=slerp(slerp(q_{i},\,q_{i+1},\,t),\,slerp(a_{i},\,a_{i+1},\,t),\,2t(1-t))
\end{eqnarray}

\begin{equation}
    a_{i}=q_{i}\exp{\left({-\frac{\log q_{i}^{*}q_{i-1}+\log q_{i}^{*}q_{i+1}}{4}}\right)}
\end{equation}

Here, the exponential of the quaternion $\exp(q)$ and the logarithm of the quaternion $\log q$ for the quaternion $q = a + bi + cj + dk$ are defined as follows:

\begin{equation}
    \exp(q)=e^{a}\left(\cos{\|bi+cj+dk\|}+\frac{bi+cj+dk}{\|bi+cj+dk\|}\sin{\|bi+cj+dk\|}\right)
\end{equation}

\begin{equation}
    \log q=\log \|q\|+\frac{bi+cj+dk}{\|bi+cj+dk\|}\arctan{\frac{\|bi+cj+dk\|}{a}}
\end{equation}

In addition, in the case of performing SQUAD-based interpolation from $q_{1}$ to $q_{2}$, and from $q_{n-1}$ to $q_{n}$, we consider $q_{0} = q_{1}$ and $q_{n + 1} = q_{n}$.


\subsubsection{Implementation}
We utilized the Kyoto Visualization System (KVS)~\cite{sakamoto2015kvs} for implementing the proposed viewpoint selection approach, based on depth and lightness entropy, as well as the sequential camera path between the selected viewpoints via SLERP- and SQUAD-based interpolation methods. KVS is a cross-platform, open-source C++ visualization library capable of running on a variety of hardware systems from traditional x86/GPU systems to GPU-less HPC systems including IBM Blue Gene L/P (PowerPC), K computer (SPARC VIIIfx), and Fugaku (ARM A64FX). KVS supports hybrid MPI/OpenMP parallelism and implements a sort-last parallel image composition method based on Binary-Swap~\cite{ma1994parallel}, with an extension to support non-power-of-two number of nodes, which is named 234Compositor~\cite{nonaka2018234compositor}. 


The pseudocode of our implementation, using the SQUAD-based interpolation,  is described in Algorithm~\ref{pseudo_in_situ}. In this pseudocode, $I []$, $V []$, and $Q []$ respectively represent the queues for storing the output image, simulation data, and the quaternion for the selected viewpoint. In addition, $is\_initial\_step(t)$ and $is\_final\_step(t)$ are functions that respectively return the \textit{true} information in the first and the final time step. The $Vis(V, \, q)$ function renders the simulation data $V$ from the viewpoint represented by the quaternion of $q$. $Entropy\_Evaluation(V)$ is the function that calculates the entropy for the simulation data $V$ at all pre-arranged viewpoints on the spherical surface, and returns the quaternion information of the viewpoint with highest entropy value. Its pseudocode is described in Algorithm~\ref{pseudo_entropy_evaluation}. In this pseudocode, $L$ represents the set of viewpoints and the $read\_back(V, \, l)$ is a function that renders the simulation data $V$ from the viewpoint $l$ and returns its frame buffer.  $entropy(f) $ is a function that calculates the entropy for the frame buffer $f$ and returns its value. $quaternion() $ is a function that returns the quaternion information from a given viewpoint.

\begin{algorithm}[t!]
\caption{In-situ visualization (using SQUAD interpolation).}
\label{pseudo_in_situ}
\begin{algorithmic}[1]
    \Function{In\_Situ\_Visualization}{$\Delta T_{V},\,\Delta T_{E}$}
        \State $I[],\,V[],\,Q[],\,t;$
        \While{$t\leq t_{end}$}
            \State $V_{t}=Sim();$
            \If{$is\_initial\_step(t)$}
                \State $Q_{t}=Entropy\_Evaluation(V_{t});$
                \State $Q.push(Q_{t});~Q.push(Q_{t});$
            \ElsIf{$t\%\Delta T_{V}==0$}
                \If{$t\%\Delta T_{E}==0$}
                    \State $Q_{t}=Entropy\_Evaluation(V_{t});$
                    \State $Q.push(Q_{t});$
                    \If{$Q.size()==4$}
                        \State $q1=Q.front();~Q.pop();$
                        \State $q2=Q.front();~Q.pop();$
                        \State $q3=Q.front();~Q.pop();$
                        \State $q4=Q.front();~Q.pop();$
                        \For{$i=0,\,1,\,\cdots,\,\Delta T_{E}-1$}
                            \State $s=i/T_{E};$
                            \State $q_{s}=squad(q1,\,q2,\,q3,\,q4,\,s);$
                            \State $V_{s}=V.front();~V.pop();$
                            \State $I_{s}=Vis(V_{s},\,q_{s});$
                            \State $I.push(I_{s});$
                        \EndFor
                    \EndIf
                \EndIf
            \EndIf
            \If{$is\_final\_step(t)$}
                \State $q1=Q.front();~Q.pop();$
                \State $q2=Q.front();~Q.pop();$
                \State $q3=Q.front();~Q.pop();$
                \State $q4=q3;$
                \For{$i=0,\,1,\,\cdots,\,\Delta T_{E}-1$}
                    \State $s=i/T_{E};$
                    \State $q_{s}=squad(q1,\,q2,\,q3,\,q4,\,s);$
                    \State $V_{s}=V.front();~V.pop();$
                    \State $I_{s}=Vis(V_{s},\,q_{s});$
                    \State $I.push(I_{s});$
                \EndFor
                \While{$V.size()>0$}
                    \State $V_{s}=V.front();~V.pop();$
                    \State $I_{s}=Vis(V_{s},\,q3);$
                    \State $I.push(I_{s});$
                \EndWhile
            \EndIf
            \State $t++;$
        \EndWhile
        \State \Return $I;$
    \EndFunction
\end{algorithmic}
\end{algorithm}

\begin{algorithm}[h]
\caption{Entropy Evaluation.}
\label{pseudo_entropy_evaluation}
\begin{algorithmic}[1]
    \Function{Entropy\_Evaluation}{$V$}
        \State $E=0.0;$
        \State $Q=1+0i+0j+0k;$
        \For{$l\in L$}
            \State $f=read\_back(V,\,l);$
            \State $e=entropy(f);$
            \If{$e>E$}
                \State $E=e;$
                \State $Q=l.quaternion();$
            \EndIf
        \EndFor
        \State \Return $Q;$
    \EndFunction
\end{algorithmic}
\end{algorithm}


\section{Experimental Evaluations}

We used the OpenFOAM CFD code and model for the experimental evaluations. The simulation model used for the evaluations was obtained from our collaborators~\cite{yoshinaga2018experimental}, and refers to a sound propagation in the oral cavity by using irregular volume data composed of 3,197,279 hexahedral elements. We integrated the in situ KVS module to the OpenFOAM code, and evaluated on two systems shown in Tables~\ref{tab:linux}~and~\ref{tab:Fugaku}. The irregular volume data was decomposed into 8 blocks for the x86 Server, and up to 1,024 blocks for the Fugaku.

\begin{table}[h]
 \caption{x86/GPU-based Server System.}
 \label{tab:linux}
 \centering
 \begin{tabular}{p{1.6cm}p{5.5cm}}
  \toprule
  {\bf Nodes} & 1 \\ 
  \addlinespace
  {\bf CPU} & Intel Xeon︎ Gold 6238R 2.20GHz 28Core$\times$2 \\
  {\bf Cores} & $28 \times 2 = 56$ \\
  {\bf RAM} & 384 GB DRAM\\
  {\bf GPU} & NVIDIA Quadro RTX8000 \\
  \addlinespace
  {\bf Compiler} & GCC version 7.5.0 \\
  {\bf MPI} & OpenMPI 2.1.1 \\
  \bottomrule
 \end{tabular}
\end{table}

\begin{table}[h]
 \caption{ARM-based Supercomputer Fugaku.}
 \label{tab:Fugaku}
 \centering
 \begin{tabular}{p{1.6cm}p{5.5cm}} 
  \toprule
  {\bf Nodes} & 158,976 \\ 
  \addlinespace
  {\bf CPU} & Fujitsu A64FX (Armv8.2-A SVE)\\
  {\bf Cores} & 48 + 2 Assistant Cores\\
  {\bf RAM} & 32GB HBM2 \\
  \addlinespace
  {\bf Compiler} & GCC-based Fujitsu Compiler Ver. 4.8.0 \\
  {\bf MPI} & OpenMPI with Fujitsu expensions for Tofu \\
  \bottomrule
 \end{tabular}
\end{table}

\subsection{Some Results}

For the initial experimental evaluations, we selected the pressure variable and used multi-isosurface rendering with three distinct isovalues that are rendered as  different colors. The total number of simulation time steps for the utilized CFD model was 15,000, and we used the parameters shown in Table~\ref{tab:parameters} for the evaluations. We evaluated the use of our proposed lightness entropy ({\it Lightness}) in addition to the depth entropy ({\it Depth}) proposed by Marsaglia~et~al.~\cite{marsaglia2021entropy}, and also the use of the average of depth and lightness entropy ({\it Depth \& Lightness}). For the use of only lightness entropy, we experimented with three diverging color maps ({\it RdBu, PiYG, PuOr}). We selected three entropy evaluation intervals ({\it 10, 30, 50}), which represent the visualization time step interval for performing the entropy calculation. We also selected three sets of viewpoints with different numbers of viewpoints in the latitude and longitude directions (latitude $\times$ longitude). Both SLERP- and SQUAD-based quaternion interpolation methods were also evaluated for estimating the camera path between the selected viewpoints. The x86/GPU Server was used for the detailed evaluation using these different parameters, and the supercomputer Fugaku was used for the scalability analysis by using up to 1024 nodes, that is, 49,152 cores in hybrid MPI/OpenMP parallelism.

\begin{table}[h]
 \caption{Parameters used for the experimental evaluations.}
 \label{tab:parameters}
 \centering
 \begin{tabular}{ll} 
  \toprule
  {\bf Entropy source} & Depth; Lightness; Depth \& Lightness \\
  {\bf Color maps} & RdBu; PiYG; PuOr \\
  {\bf \# of viewpoints} & $15 \times 30$; $25 \times 50$; $35 \times 70$ \\
  {\bf Intervals} ($N_{E}$) & 10; 30; 50 \\
  {\bf Interpolation} & SLERP; SQUAD \\
  \bottomrule
 \end{tabular}
\end{table}



\begin{figure*}[htb]
    \centering
    \includegraphics[width=0.85\linewidth]{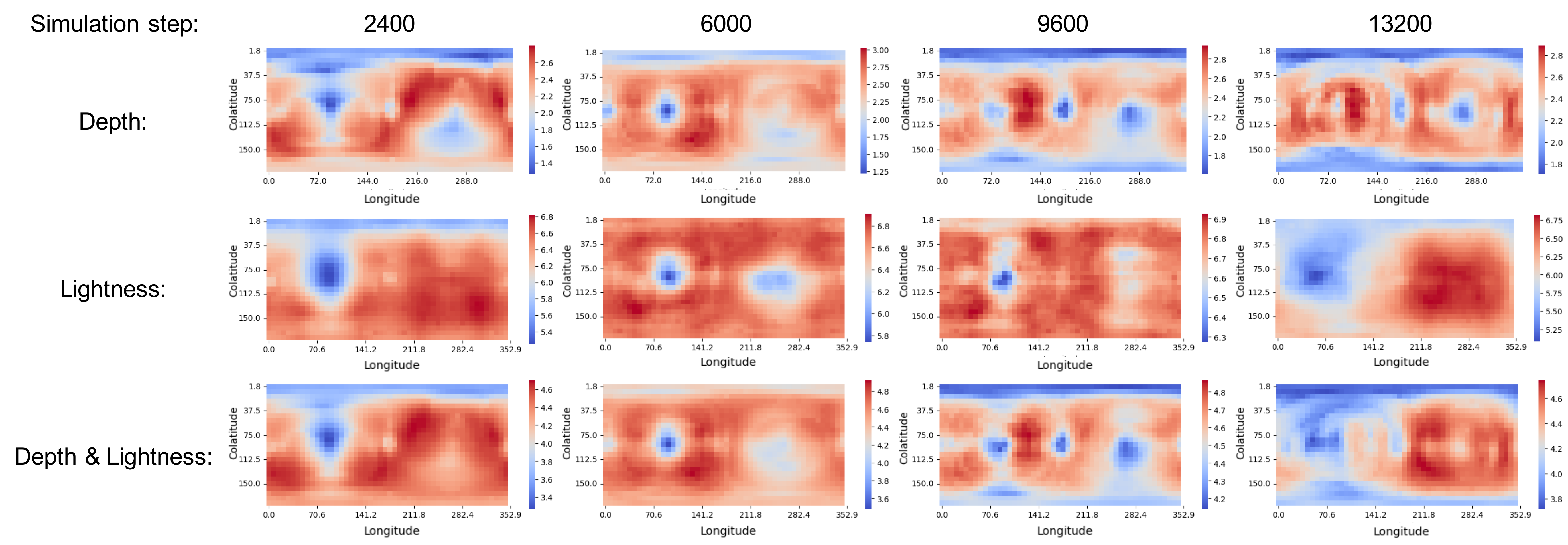}
    \caption{Entropy heatmaps for different entropy sources.}
    \label{fig:heatmap_entropy_sources}
\end{figure*}

\begin{figure*}[htb]
    \centering
    \includegraphics[width=0.85\linewidth]{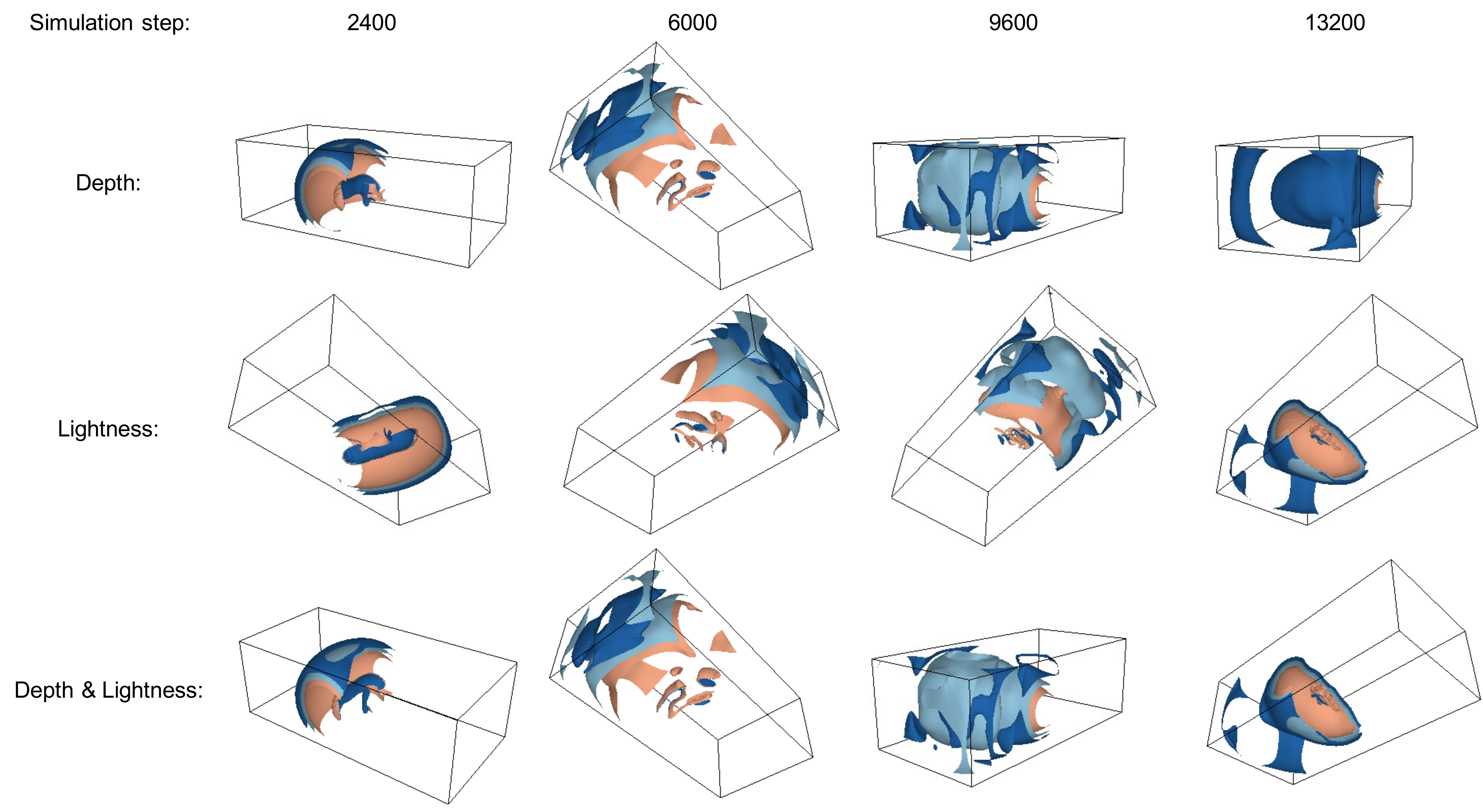}
    \caption{Rendered images from the selected viewpoints.}
    \label{fig:rendering_entropy_sources}
\end{figure*}

\begin{figure*}[htb]
    \centering
    \includegraphics[width=0.85\linewidth]{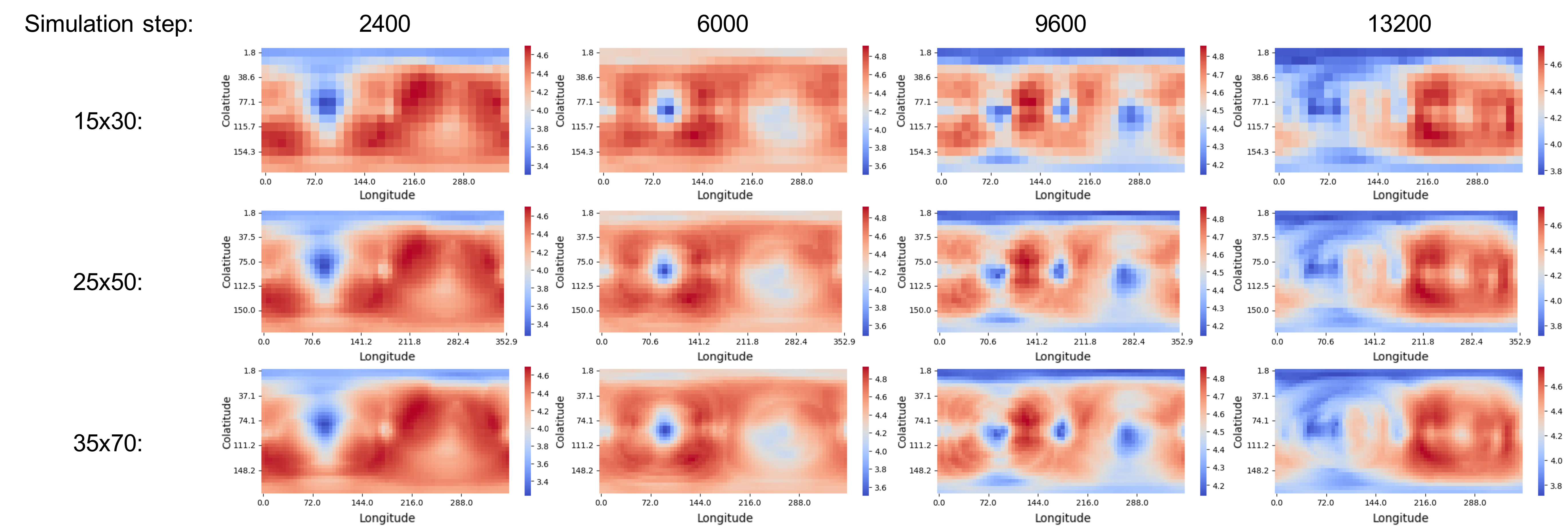}
    \caption{Lightness Entropy heatmaps for different diverging color maps.}
    \label{fig:heatmap_colormaps}
\end{figure*}

\begin{figure*}[htb]
    \centering
    \includegraphics[width=0.85\linewidth]{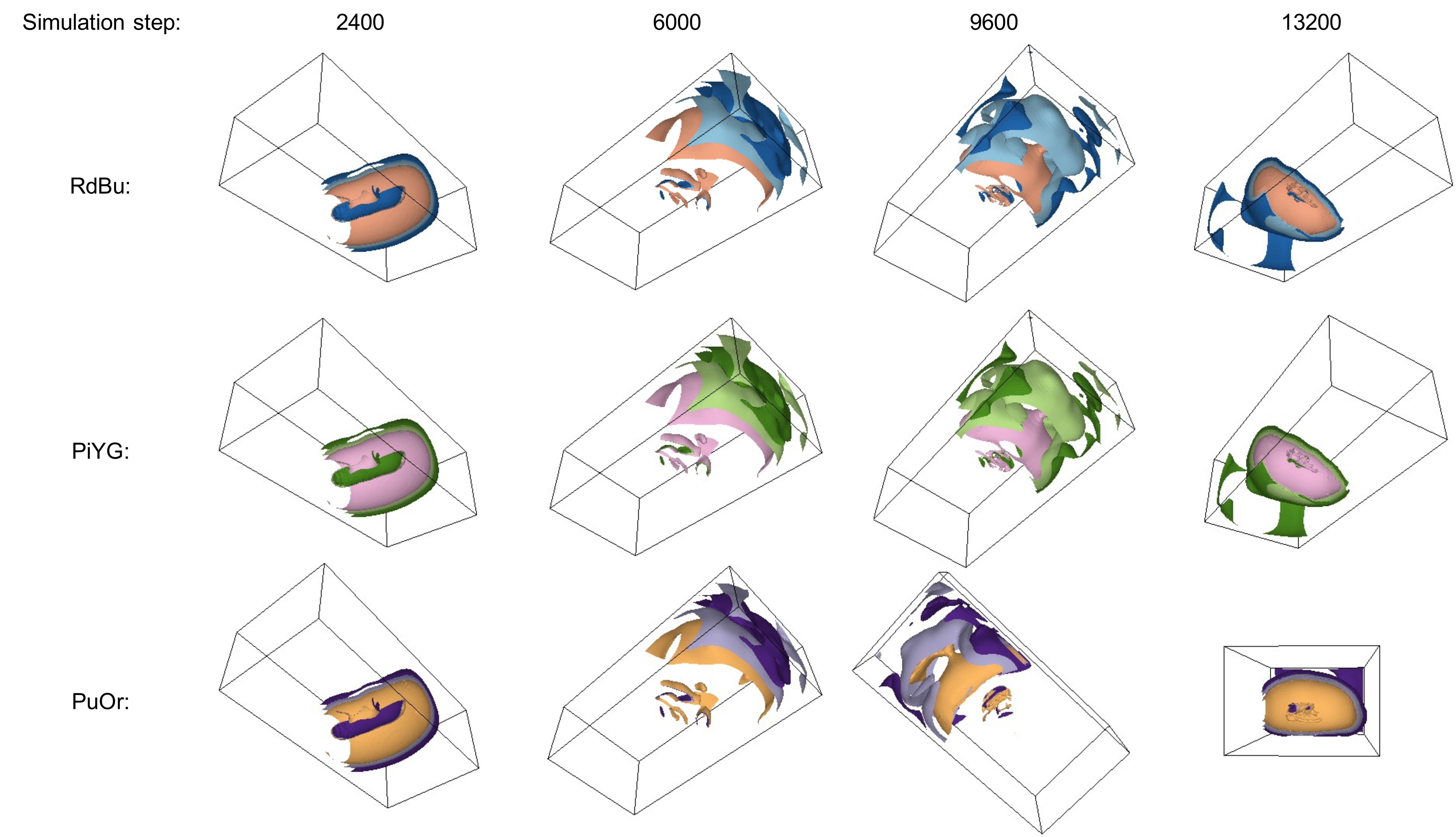}
    \caption{Rendered images from the selected viewpoints for different diverging color maps.}
    \label{fig:rendering_colormaps}
\end{figure*}

\begin{figure*}[htb]
    \centering
    \includegraphics[width=0.85\linewidth]{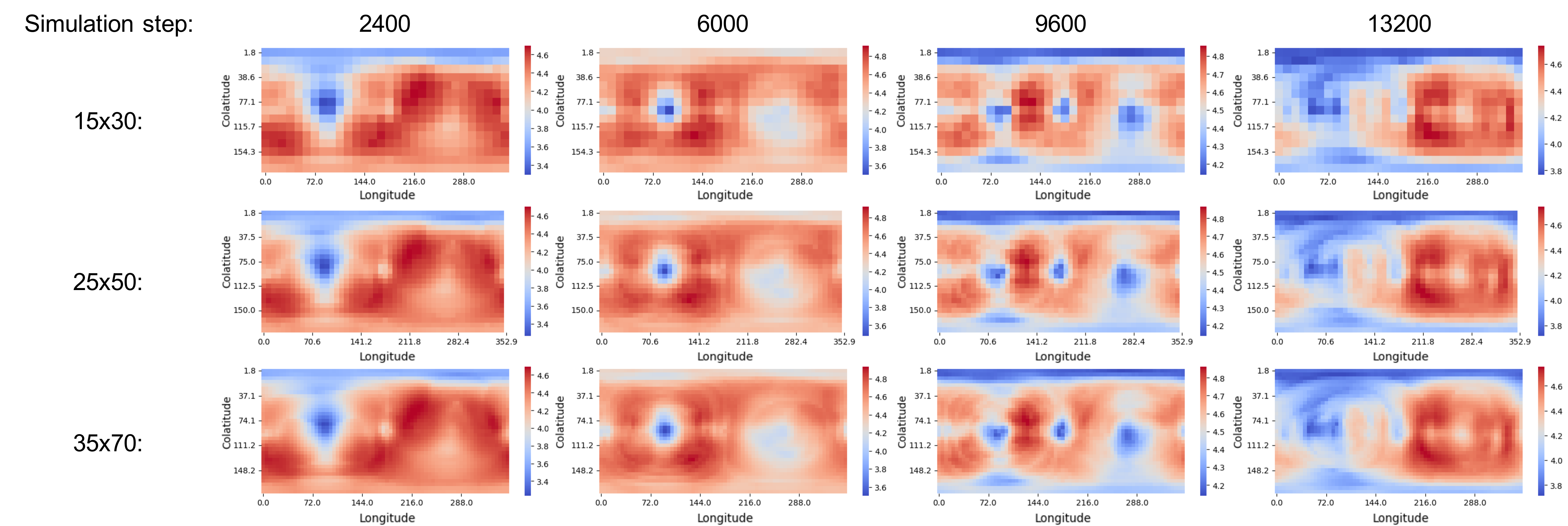}
    \caption{Entropy heatmaps when varying the number of viewpoints.}
    \label{fig:heatmap_number_viewpoints}
\end{figure*}

\begin{figure*}[htb]
    \centering
    \includegraphics[width=0.85\linewidth]{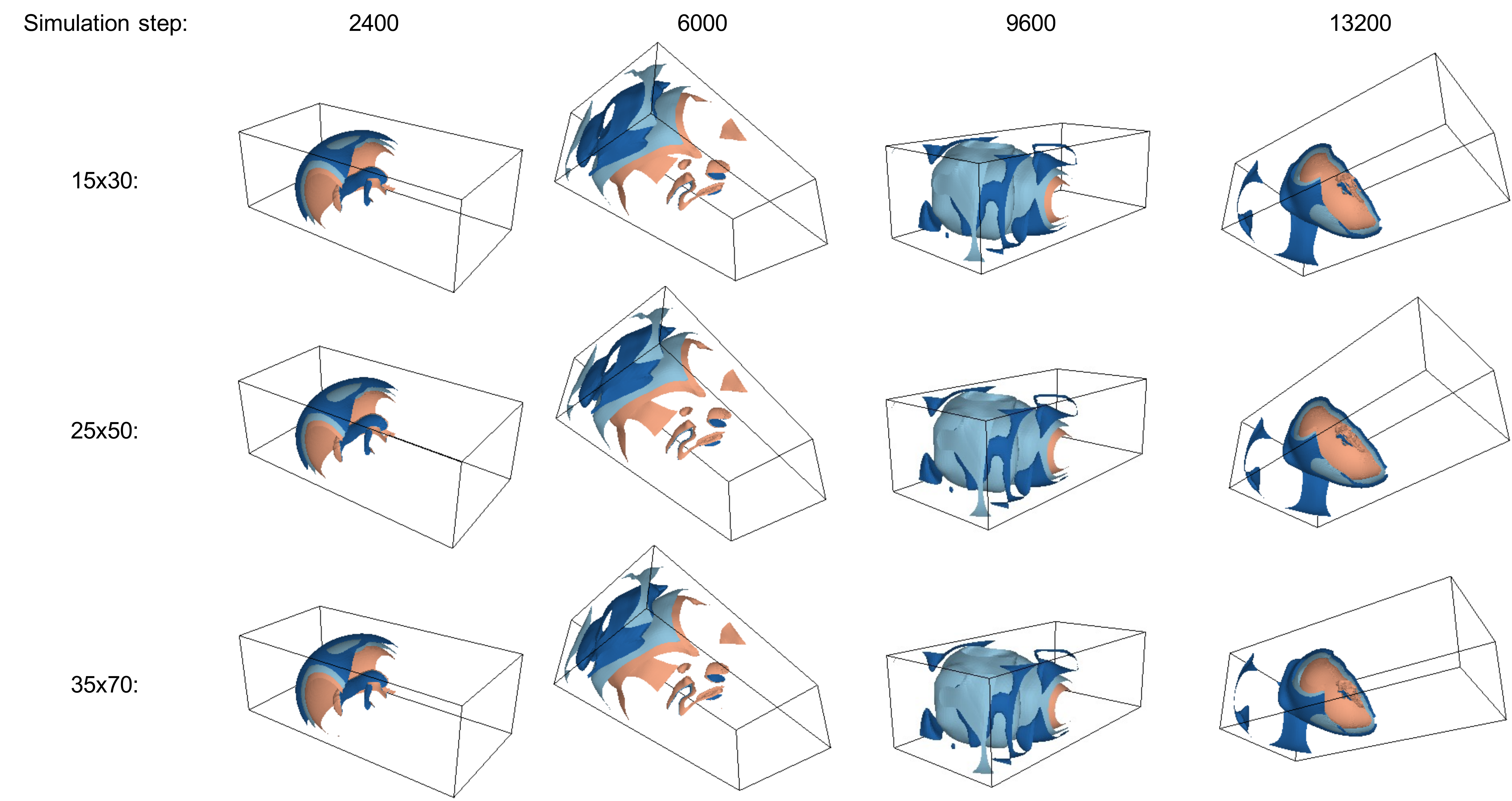}
    \caption{Rendered images from the selected viewpoints when varying the number of viewpoints.}
    \label{fig:rendering_number_viewpoints}
\end{figure*}

Fig.~\ref{fig:heatmap_entropy_sources} shows some entropy heatmaps evaluated by using \color{black}all three  entropy sources at different simulation time steps (2400, 6000, 9600, and 13200). \color{black}The set of viewpoints evenly distributed on the spherical surface is mapped onto the 2D heatmap where the viewpoints on the same latitude are placed on the same horizontal axis, and in the same manner, the viewpoints on the same longitude are placed on the same vertical axis. The blue-colored regions show the portions where the evaluated entropy has low value, and on the other hand, the red-colored regions show the portions where the evaluated entropy has high value. Fig.~\ref{fig:rendering_entropy_sources} shows the multi-isosurface rendered results from the selected viewpoints obtained in Fig.~\ref{fig:heatmap_entropy_sources}; Fig.~\ref{fig:rendering_colormaps} shows the multi-isosurface rendered results from the selected viewpoints obtained in Fig.~\ref{fig:heatmap_colormaps}; Fig.~\ref{fig:rendering_number_viewpoints} shows the multi-isosurface rendered results from the selected viewpoints obtained in Fig.~\ref{fig:heatmap_number_viewpoints}. \color{black}Table~\ref{tab:average_time_metrics} shows the average elapsed time of entropy calculation per image for different entropy sources when using an image size of $512 \times 512$ on the x86/GPU-based Server System. Compared to depth entropy, we can observe that the computational costs when using lightness become much higher. In addition, we can verify that the number of viewpoints directly influences the computational cost. Code optimizations and the use of parallel processing for trying to reduce this computational cost are planned as future works.\color{black} Table~\ref{tab:average_entropy_viewpoints} shows a comparison of output images' average entropy when varying the number of viewpoints. Here, the utilized entropy source is Depth \& Lightness, the entropy evaluation interval is 30, and the interpolation method is SQUAD. \color{black}We can observe that the difference in the average entropy when varying the number of viewpoints is small.\color{black}


\begin{table}[htb!]
 \caption{Average elapsed time of entropy calculation for different entropy sources (x86 System).}
 \label{tab:average_time_metrics}
 \centering
 \begin{tabular}{cc} 
  \toprule
  {\bf Entropy Sources} & {\bf Average elapsed time [s]} \\
  \midrule
  Depth & 2.24e-4 \\
  Lightness & 1.30e-3 \\
  Depth \& Lightness & 1.53e-3 \\
  \bottomrule
 \end{tabular}
\end{table}

\begin{table}[htb!]
 \caption{Average entropy when varying the number of viewpoints.}
 \label{tab:average_entropy_viewpoints}
 \centering
 \begin{tabular}{cc} 
  \toprule
  {\bf \# of viewpoints} & {\bf Average entropy} \\
  \midrule
  $15 \times 30$ & 3.09 \\
  $25 \times 50$ & 3.10 \\
  $35 \times 50$ & 3.09 \\
  \bottomrule
 \end{tabular}
\end{table}

\begin{figure}[htb!]
    \centering
    \includegraphics[width=7.5cm]{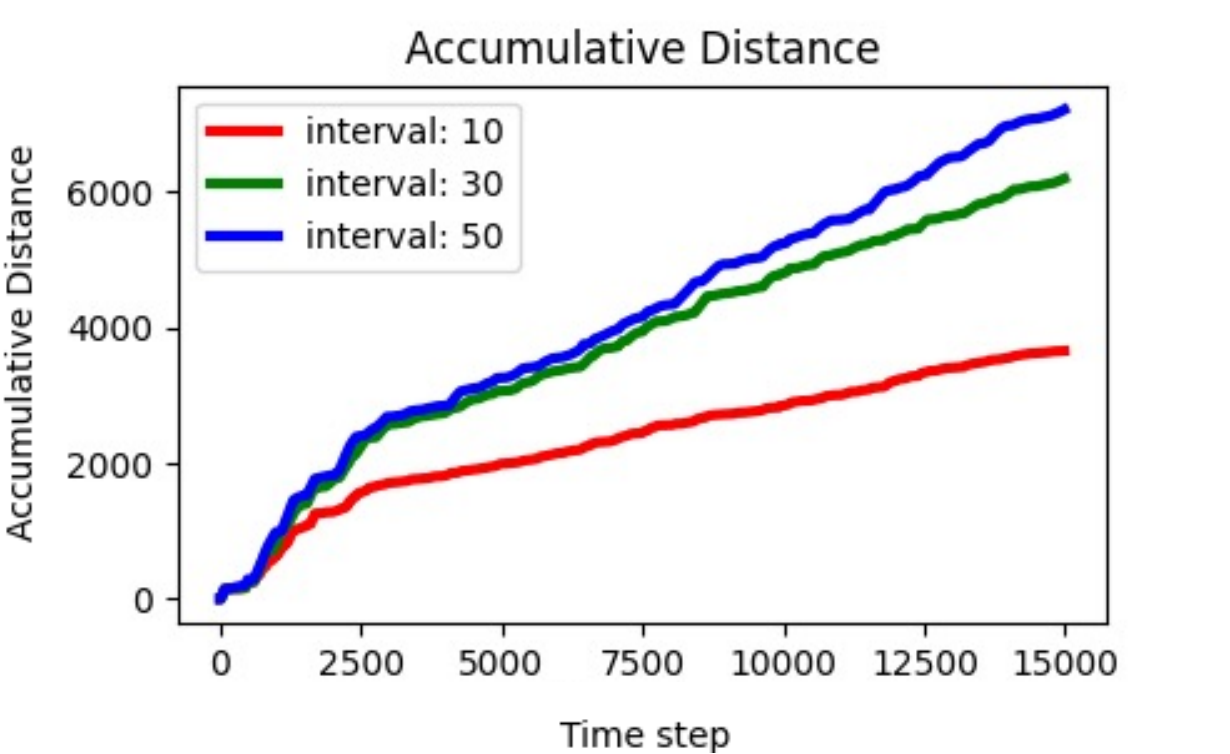}
    \caption{Accumulative distance from the interpolated camera path to the viewpoints, with the highest entropy, at each visualization time step, for different entropy evaluation intervals.}
    \label{fig:accumulative_distance_interval}
\end{figure}
\begin{figure}[htb]
    \centering
    \includegraphics[width=7.5cm]{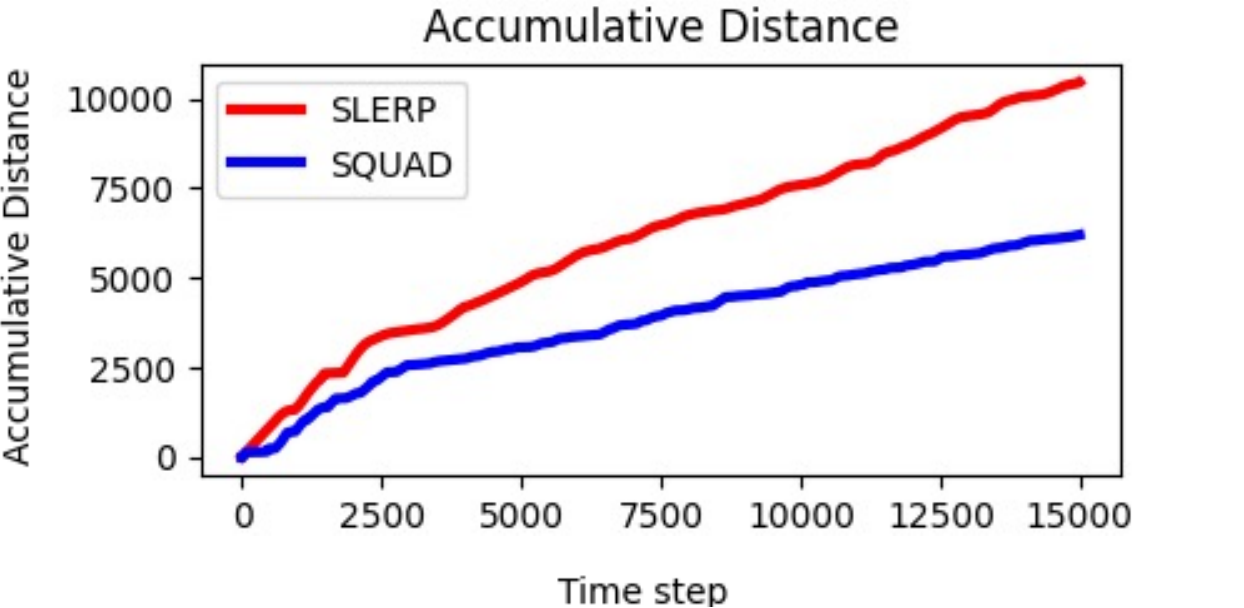}
    \caption{Accumulative distance to the viewpoints, with the highest entropy, at each visualization time step, for different interpolation methods.}
    \label{fig:accumulative_distance_interpolation}
\end{figure}

Fig.~\ref{fig:accumulative_distance_interval} shows a comparison of the accumulative distance from the estimated camera path position to the viewpoint with the highest entropy, at each visualization time step, for different entropy evaluation intervals\color{black}; Fig.~\ref{fig:accumulative_distance_interpolation} shows a comparison of the accumulative distance for different interpolation methods.\color{black} Table~\ref{tab:average_entropy_interval} shows a comparison of output images' average entropy for different entropy evaluation intervals. Here, the utilized entropy source is Depth \& Lightness, the number of viewpoints is $25 \times 50$, and the interpolation method is SQUAD. In the case of $N_{E}=1$, rendered images of the viewpoint with the highest entropy at each visualization time step are output. From this figure and table, we can observe that as the entropy evaluation interval increases, the accumulative distance also increases, and the amount of average entropy decreases. \color{black}Table~\ref{tab:average_time_interpolations} shows the average elapsed time for path calculation between two selected viewpoints for different interpolation methods. We can observe that the computational cost is proportional to the number of intervals, and the cost of SQUAD is much higher than that of SLERP. However, it is worth noting that the influence on the total computational cost compared to the entropy calculation cost is small and almost neglectable. \color{black} Table~\ref{tab:average_entropy_interpolation} shows a comparison of output images' average entropy for different interpolation methods. Here, the entropy source is Depth \& Lightness, the number of viewpoints is $25 \times 50$, and the entropy evaluation interval is 30. We can observe that when selecting SQUAD, the accumulative distance becomes smaller and achieves a slight increase in average entropy. 

\begin{table}[htb!]
 \caption{Average entropy for different entropy evaluation intervals.}
 \label{tab:average_entropy_interval}
 \centering
 \begin{tabular}{cc} 
  \toprule
  {\bf Intervals} ($N_{E}$) & {\bf Average entropy} \\
  \midrule
  1 & 3.17 \\
  10 & 3.12 \\
  30 & 3.10 \\
  50 & 3.08 \\
  \bottomrule
 \end{tabular}
\end{table}

\begin{table}[htb!]
 \caption{Average elapsed time for path calculation between two selected viewpoints using different interpolation methods (x86 System).}
 \label{tab:average_time_interpolations}
 \centering
 \begin{tabular}{cccc}
  \toprule
  {\bf Interpolation} & \multicolumn{3}{c}{\bf Intervals ($N_E$)} \\
  \cline{2 - 4}
  {\bf method} & 10 & 30 & 50 \\
  \midrule
  SLERP & 2.80e-6 & 5.16e-6 & 6.93e-6\\
  SQUAD & 9.80e-6 & 3.04e-5 & 3.54e-5\\
  \bottomrule
 \end{tabular}
\end{table}

\begin{table}[htb!]
 \caption{Average entropy for different interpolation methods.}
 \label{tab:average_entropy_interpolation}
 \centering
 \begin{tabular}{cc} 
  \toprule
  {\bf Interpolation method} & {\bf Average entropy} \\
  \midrule
  SLERP & 3.09 \\
  SQUAD & 3.10 \\
  \bottomrule
 \end{tabular}
\end{table}

\subsection{Discussions}



Regarding the influence of different entropy sources \color{black}(Fig.~\ref{fig:heatmap_entropy_sources}), we observed that the lightness entropy has a higher influence than the depth entropy, and has an even higher influence when using both depth and lightness entropy. Therefore, we opted to add both depth and lightness entropies after normalization. In addition, due to a large number of viewpoints with high entropy, the sequentially selected viewpoints can be separated far apart from each other thus resulting in an intense camera movement over the entire visualization time steps. We also observed that when using the lightness entropy, the entropy calculation took a little more time than using the depth entropy. This was because of the necessary conversion from RGB values to lightness values. \color{black}It is worth noting that the selection of the viewpoint evaluation metric will depend on the targeted simulation, visualization method, and users' analysis goals. Therefore, to satisfy {\bf R1}, it becomes important to implement a variety of viewpoint evaluation metrics to handle different use case combinations. In addition, depending on the use case, it may be helpful that different evaluation metrics are interchangeable at run time in an adaptive manner.

Regarding the influence of the diverging color maps for the lightness entropy, we initially perceived almost no difference between the heatmaps. \color{black}However, there was a slight difference among them, and at certain time steps, we observed that the selected viewpoints were also different. Among the color maps, heatmaps for the PuOr was especially different from the others. This may be because the change in the lightness of the PuOr was also different from the other diverging color maps.




\color{black}Regarding the influence of the entropy evaluation intervals, as this interval becomes smaller there will be fewer complementary images between the selected viewpoints. As a result, changes in viewpoint may become intense in a short period of time, this will lead to a non-smooth video which affects the users' post-hoc visual analysis tasks. It is worth noting that when the simulation state is not expected to change rapidly, there will be no necessity to frequently evaluate the viewpoints. However, when utilizing larger entropy evaluation intervals, a larger amount of memory will be required for temporarily storing the simulation data. That is, there is a trade-off between the entropy evaluation intervals and the memory consumption, and as a result, depending on the simulation time step range and simulation data size, large entropy evaluation intervals, such as the utilized $N_{E}=30$ and $N_{E}=50$, may be sufficient to satisfy the {\bf R2}. 

Regarding the influence of the number of viewpoints on the spherical surface, we verify that there was no significant difference for varying number of viewpoints. However, it is worth noting that the computational time required to select the viewpoints will increase proportionately with the increase in the number of viewpoints.

Regarding the influence of the quaternion interpolation method for estimating the camera path between selected viewpoints, we observed that the camera path using SQUAD-based interpolation passes closer to the viewpoint with highest entropy at the intermediate time steps. We also observed that jerky camera movements tend to occur when using the SLERP-based interpolation. On the other hand, smoother camera movement was observed when using the SQUAD-based interpolation, and as a result, we can consider that it will cause less discomfort to the user when seeing the animated rendering results since the camera movement will be more natural. Therefore, we can consider that SQUAD-based quaternion interpolation satisfies the {\bf R2}.

Moreover, we carried out some evaluations with the domain scientists who \color{black} assisted in the development of previous work on in-situ adaptive timestep selection~\cite{yamaoka2019situ}. We obtained technical feedback from the generated visualization results in the form of animated videos. \color{black} According to them, the video generated by using the proposed method seems to present more information than the video generated by using fixed viewpoint camera settings, \color{black} which has traditionally been used in their simulation analysis. However, they also pointed out that the proposed video gives the impression of excessive movement and sometimes tracking phenomena that do not need much attention. As some suggestions, they mentioned that it would be better to slightly reduce high viewpoint variations or suppress unnecessary movement, and to improve evaluation methods for the viewpoint selection. \color{black}As an additional suggestion, they would prefer to have the ability to zoom in on the target object to enable closer observation. \color{black}These suggestions will be taken into consideration for further developments planned as future works.\color{black}


\color{black}In our current implementation, the set of volume data in the entropy evaluation interval needs to be stored in the memory before the processing, and this memory cost can become an impediment for memory-hungry simulations. However, we consider that this approach can be useful during test runs and model calibration runs, before the main simulation run, when smaller models are usually sufficient. In addition, the in-transit approach for flushing the simulation data from the memory to another node or even system can be considered helpful for minimizing this problem and is planned for future work. Another planned future work is the application of the adaptive timestep sampling~\cite{yamaoka2019situ} where larger time intervals will be assigned to timestep regions with small variations between the simulation results. This larger entropy evaluation time step by skipping some simulation results may be helpful for accelerating the visualization processing as well as reducing the excessive movements pointed out by the domain scientists.\color{black}



\section{Conclusions}






In this work, we proposed an information entropy-based camera path estimation method for in-situ visualization. Considering that most of the images generated by traditional batch-based tightly coupled in-situ visualization may have small or even no contribution for the post-hoc visual analysis, we focused on generating a smooth video that tries to provide as much information as possible to facilitate the rapid understanding of the simulation or \color{black}to narrow down the spatio-temporal region of interest for posterior detailed analysis such as by using traditional image-based visualization. \color{black}The proposed method focuses on selecting the most appropriate viewpoints, based on information entropy, at regular intervals. Intermediate images are generated from the estimated camera path connecting these selected viewpoints, and the produced smooth video that is produced is expected to be helpful for understanding the underlying simulation phenomena. From the experimental evaluations and feedback from domain scientists, we can confirm that the video generated by the proposed approach provides more information compared to those generated by using fixed viewpoint camera settings. However, there is still need for improvements, and we can cite the following targets for future works: implementation of better evaluation methods for the viewpoint selection; implementation of zoom in and out functionalities; \color{black}integration with the adaptive timestep sampling (irregular time intervals); improvement of computational performance such as by applying parallel processing; \color{black} and estimation of the focal point for the camera.



\acknowledgments{
The authors are grateful to Tsukasa Yoshinaga (Toyohashi University of Technology) and Kazunori Nozaki (Osaka University) for the simulation model and technical feedback. This work was partially supported by JSPS KAKENHI (Grant Numbers: 20H04194, 21H04903, 22H03603), and the National Key R\&D Program of China under Grant No. 2021YFE0108400. This work used computational resources of supercomputer Fugaku provided by the RIKEN Center for Computational Science.}

\bibliographystyle{abbrv-doi}

\bibliography{2023_PacificVis__KenIwata}
\end{document}